\RequirePackage{arydshln}
\documentclass[aps,twocolumn,nofootinbib,superscriptaddress,preprintnumbers,pra,10pt,floatfix]{revtex4-1}

\usepackage[utf8]{inputenc}

\usepackage{soul}

\usepackage{float}
\usepackage{amsmath,amssymb}
\usepackage{dsfont} 
\usepackage{hyperref}
\usepackage{graphicx}
\usepackage{enumitem}
\usepackage{mathtools}
\usepackage{bbold}
\usepackage{multirow}
\usepackage{ytableau}
\usepackage{youngtab}
\usepackage{braket}
\usepackage{soul} 
\usepackage{notes2bib}

\usepackage{colortbl} 
\definecolor{Gray}{gray}{0.95}
\definecolor{RGray}{gray}{0.90}
\definecolor{CGray}{gray}{0.92}

\newcommand{\eminus}{\vcenter{\hbox{\scalebox{0.75}[1]{$ - $}}}}

\usepackage{arydshln}

\makeatletter
\g@addto@macro\bfseries{\boldmath}
\makeatother

\makeatletter
\renewcommand\paragraph{\@startsection{paragraph}{4}{\z@}%
                                    {3.25ex \@plus1ex \@minus.2ex}%
                                    {-1em}%
                                    {\normalfont\normalsize\bfseries}}
\makeatother

\allowdisplaybreaks
\begin{document}

\preprint{ZU-TH-08/22}

\title{Flavor hierarchies, flavor anomalies, and Higgs mass from a warped extra dimension}

\author{Javier Fuentes-Mart\'{\i}n}
\email{javier.fuentes@ugr.es}
\affiliation{Departamento de Física Teórica y del Cosmos, Universidad de Granada, E–18071 Granada, Spain}
\author{Gino Isidori}
\email{isidori@physik.uzh.ch}
\affiliation{Physik-Institut, Universit\"at Z\"urich, CH-8057 Z\"urich, Switzerland}
\author{Javier M. Lizana}
\email{jlizana@physik.uzh.ch}
\affiliation{Physik-Institut, Universit\"at Z\"urich, CH-8057 Z\"urich, Switzerland}
\author{Nud{\v z}eim Selimovi{\'c}}
\email{nudzeim@physik.uzh.ch}
\affiliation{Physik-Institut, Universit\"at Z\"urich, CH-8057 Z\"urich, Switzerland}
\author{Ben A. Stefanek}
\email{bestef@physik.uzh.ch}
\affiliation{Physik-Institut, Universit\"at Z\"urich, CH-8057 Z\"urich, Switzerland}

\begin{abstract}
The recent $B$-meson anomalies are coherently explained at the TeV scale by \textit{4321 gauge models} with hierarchical couplings reminiscent of the Standard Model Yukawas. We show that such models arise as the low-energy limit of a complete theory of flavor, based on a warped fifth dimension where each Standard Model family is quasi-localized in a different brane. The Higgs is identified as a pseudo-Nambu-Goldstone boson emerging from the same dynamics responsible for 4321 symmetry breaking. This novel construction unifies quarks and leptons in a flavor non-universal manner, provides a natural description of flavor hierarchies, and addresses the electroweak hierarchy problem.
\end{abstract}

\maketitle

\allowdisplaybreaks

\section{Introduction}\label{sec:intro}

The deviations from lepton flavor universality observed in neutral-current~\cite{LHCb:2014vgu,LHCb:2017avl,LHCb:2019hip,LHCb:2021trn} and charged-current~\cite{BaBar:2012obs,BaBar:2013mob,Belle:2015qfa,LHCb:2015gmp,LHCb:2017smo,LHCb:2017rln} semileptonic $B$~decays have stimulated intense  model-building activity, triggering new ideas about the ultraviolet (UV) completion of the Standard Model (SM). Two key aspects have emerged quite clearly from the early attempts to provide a combined explanation of the two sets of anomalies: i)~a possible common origin of flavor anomalies and Yukawa hierarchies~\cite{Bordone:2017bld}, as hinted by the approximate $U(2)^n$ flavor structure of new physics ~\cite{Greljo:2015mma,Barbieri:2015yvd}, ii)~the necessity  of new degrees of freedom at the TeV scale coupled mainly to the SM third generation, hinting at a possible link with the electroweak (EW) hierarchy problem~\cite{Barbieri:2017tuq,Fuentes-Martin:2020bnh}. 

In this letter, we show how these two aspects can be consistently combined within a five-dimensional (5D) model. The three main assumptions of our construction, and their motivations, can be listed as follows:

\smallskip
{\bf I.} {\em 4321 gauge symmetry above the TeV scale}.  The most effective mediator to address both sets of anomalies is a TeV-scale $U_1$ leptoquark~\cite{Alonso:2015sja,Calibbi:2015kma,Barbieri:2015yvd,Bhattacharya:2016mcc,Buttazzo:2017ixm}. This field can be identified with one of the broken generators of a fundamental $4321$ gauge group, where color is the diagonal subgroup of $SU(4)_h \times SU(3)_l$~\cite{DiLuzio:2017vat}. The  labels $h$ and $l$ indicate the flavor non-universal assignment of the SM fermions under this part of the gauge group, resulting in a $U_1$ coupled mainly to third-generation fermions~\cite{Bordone:2017bld,Greljo:2018tuh}. Apart from the $U_1$, two EW-neutral gauge bosons acquire mass from the 4321 breaking: a color-octet, $G^\prime$, and a singlet, $Z^\prime$. The presence of these two mediators do not alter the $U_1$ solution of the anomalies~\cite{DiLuzio:2018zxy,Cornella:2019hct,Cornella:2021sby}.

\smallskip
{\bf II.} {\em Flavor hierarchies from a 3-brane structure in 5D}. The hierarchies in both the Yukawa and $U_1$ couplings, i.e.~the breaking of the approximate $U(2)^n$ flavor symmetry acting on the light families at the TeV scale, emerge from a multi-scale construction~\cite{Panico:2016ull,Bordone:2017bld,Allwicher:2020esa,Barbieri:2021wrc}
that, in turn, can be viewed as the effect of a 3-brane structure in 5D. The strong constraints on flavor-violating terms involving the light families naturally point toward a warped geometry~\cite{Fuentes-Martin:2020pww}.
The size of the Yukawa couplings implies $kL \approx 10$~\cite{Fuentes-Martin:2020pww}, where $L$ is the distance between the infrared (IR) (3$^{\rm rd}$ gen.) brane and the most UV (1$^{\rm st}$ gen.) brane, and $k$ is the 5D curvature constant. 

\smallskip
{\bf III.} {\em Holographic Higgs}.  
The SM Higgs can be realized as a pseudo-Nambu-Goldstone boson (pNGB) emerging from the same dynamics responsible for the breaking of $SU(4)_h \times SU(3)_l$~\cite{Fuentes-Martin:2020bnh}. In the warped 5D description, this can be achieved via gauge-Higgs unification~\cite{Contino:2003ve}, realized by extending the EW part of the bulk gauge symmetry. For the sake of simplicity and minimality, we assume 
\begin{align}\label{eq:GbulkIR}
\begin{aligned}
\mathcal{G}^{\rm 23}_{{\rm bulk}} &\equiv SU(4)_h\times SU(3)_l\times U(1)_l \times  SO(5)\,, \\
\mathcal{G}_{\rm IR} &\equiv SU(3)_c\times U(1)_{B-L}\times SO(4)\,,
\end{aligned}
\end{align} 
where $SU(3)_c$ and $U(1)_{B-L}$ are flavor-universal 
subgroups of $SU(4)_h\times SU(3)_l \times U(1)_l$, and the 23 bulk 
is the most IR side of the bulk (see Fig.~\ref{fig:branes}). 
The fifth component of the gauge fields associated to the $SO(5)/SO(4)$ coset contains four pNGB zero modes transforming as a ${\bf 4}$ of $SO(4)$ that we identify with the SM Higgs field, thus realizing the minimal composite Higgs scenario~\cite{Agashe:2004rs}.

\section{The 5D model}\label{sec:model}

We consider a 5D model with a warped compact extra dimension containing three branes (similar to the one explored in~\cite{Kogan:2000xc}) with two positive and one negative tension branes (++$-$).  The metric is
\begin{align}
ds^{2} = e^{-2\sigma(y)}\, \eta_{\mu\nu}\, dx^{\mu} dx^{\nu} - dy^{2} \,,
\end{align}
where, in the absence of backreaction from scalar fields, the warp factor $\sigma(y)$ is  
\begin{align}\label{eq:warp}
\sigma(y) = \begin{cases} 
      \sigma_1(y)=k_{1} y & 0\leq y \leq \ell \\
      \sigma_2(y)=k_{1} \ell + k_{2}(y-\ell) & \ell \leq y \leq L
   \end{cases} \,.
\end{align}
Here $L$ denotes the total length of the compact extra dimension and $\ell$ the location of the intermediate brane. Due to the tension of the branes, $k_1\leq k_2$.
For simplicity, in what follows we assume $k_{1} = k_{2} \equiv k$, corresponding to a zero middle brane tension. This multi-brane setup can be stabilized via a straight-forward extension of the Goldberger-Wise (GW) mechanism~\cite{Goldberger:1999uk,Lee:2021wau}, with suitable brane-localized potentials for the GW scalar~\bibnote{In the limit of $k_{1}\ell, k_{2}(L-\ell) > 1$ and small bulk mass for the GW scalar, the stabilization is achieved via two factorized solutions of the type found in~\cite{Goldberger:1999uk}, one for each distance}. 

Beside the IR bulk and brane symmetries specified in \eqref{eq:GbulkIR}, we assume
\begin{align}
\mathcal{G}^{\rm 12}_{{\rm bulk}} &\equiv SU(4)_h\times SU(4)_l\times SO(5)\,, \\  
\mathcal{G}_{\rm UV} & \equiv SU(4)_h\times SU(3)_l\times U(1)_l\times SU(2)_L\times U(1)_R\,, \nonumber
\end{align}
where $\mathcal{G}_{\rm UV}$ and $\mathcal{G}^{\rm 12}_{{\rm bulk}}$ are the gauge symmetries in the most UV brane and the UV side of the bulk, respectively. The middle brane is therefore a discontinuity corresponding to the symmetry-breaking pattern $SU(4)_l \rightarrow SU(3)_l \times U(1)_l$. Alternatively, we could have chosen $\mathcal{G}^{\rm 12}_{{\rm bulk}}=\mathcal{G}^{\rm 23}_{{\rm bulk}}$, with no difference in the low-energy phenomenology. In this case, light-family quark-lepton unification could take place in a bulk between a deeper UV brane (e.g. the Planck brane) and the first-family brane.


The chosen gauge symmetries yield $15+4$ pNGBs, 15 of which become the longitudinal components of the 4321 gauge bosons, $U_1$, $G'$ and $Z'$, which acquire a degenerate mass of $M_{15} \approx \Lambda_{\rm IR} \sqrt{2/(kL)}$~\bibnote{Assuming no boundary kinetic terms for $SU(4)_h\times SU(3)_\ell \times U(1)_{\ell+R}$ part of the gauge symmetry}, with $\Lambda_{\rm IR}\approx k \, e^{-kL}$. This mass generation mechanism (similar to the one in~\cite{Fuentes-Martin:2020bnh}), yields a mass gap between the 4321 gauge bosons and the lightest vector resonances of the Kaluza-Klein (KK) tower, namely $M_{\rm KK}/M_{15} \approx \sqrt{2kL}$. The remaining pNGBs correspond to the SM Higgs field: $H\sim(\mathbf{1},\mathbf{2})_{1/2}$.

\begin{table}[t]
    \renewcommand{\arraystretch}{1.2}
    \centering
    \begin{tabular}{|c|ccc||cc|}
    \hline
    Field & $SU(4)_h$ & $SU(4)_l$ & $SO(5)$ & $U(1)_\Psi$ & $U(1)_\mathcal{S}$ \\
    \hline
    \hline
    $\Psi^3,\Psi_d^3,\mathcal{X}^{({\prime})}$ & $\mathbf{4}$ & $\mathbf{1}$ & $\mathbf{4}$ & $1$ & $0$\\
    $\Psi^j,\Psi_{u,d}^j$        & $\mathbf{1}$ & $\mathbf{4}$ & $\mathbf{4}$ & $1$ & $0$\\
    $\mathcal{S}^i$         & $\mathbf{1}$ & $\mathbf{1}$ & $\mathbf{1}$ & $0$ & $1$\\
    \hline
    \hline
    $\Sigma$ & $\mathbf{1}$ & $\mathbf{1}$ & $\mathbf{5}$ & $0$ & $0$\\
    $\Omega$ & $\mathbf{1}$ & $\mathbf{4}$ & $\mathbf{4}$ & $1$ & $-1$\\ 
    $\Phi$   & $\mathbf{1}$ & $\mathbf{1}$ & $\mathbf{1}$ & $0$ & $2$\\ 
    \hline
    \end{tabular}
    \caption{Matter content. Here, $i=1,2,3$ and $j=1,2$. The upper block refers to fermion fields, the lower block to scalars.}
    \label{tab:content}
\end{table}

The matter fields and their corresponding transformation properties under the 12-bulk gauge symmetry are listed in Table~\ref{tab:content}. We embed all fermion fields (except $\mathcal{S}^i$) into the spinorial $\mathbf{4}$ representation of $SO(5)$, which contains two (complex) doublets, one transforming in the fundamental of $SU(2)_L$ and the other in the fundamental of $SU(2)_R$. These fermions are also embedded into fundamental representations of $SU(4)_{h,l}$ forming quark-lepton multiplets \`a la Pati-Salam. The boundary conditions (BCs) for the fermions are chosen as follows
\begin{align}\label{eq:FermionBCs}
\Psi^3&=
\begin{bmatrix}
\psi^3\,(+,+)\\[2pt]
\psi_u^3\,(-,-)\\[2pt]
\tilde\psi_d^3\,(+,-)\\
\end{bmatrix}
\,,&
\Psi^3_d&=
\begin{bmatrix}
\tilde\psi^3\,(+,-)\\[2pt]
\tilde\psi_u^3\,(+,-)\\[2pt]
\psi_d^3\,(-,-)\\
\end{bmatrix}
\,,\nonumber\\
\mathcal{X}^{(\prime)}&=
\begin{bmatrix}
\chi^{(\prime)} (\pm,\pm)\\[2pt]
\chi^{(\prime)}_u\,(\mp,\pm)\\[2pt]
\chi^{(\prime)}_d\,(\mp,\pm)\\
\end{bmatrix}
\,,&
\Psi^j&=
\begin{bmatrix}
\psi^j\,(+,+)\\[2pt]
\tilde\psi_u^j\,(-,+)\\[2pt]
\tilde\psi_d^j\,(-,+)\\
\end{bmatrix}
\,,\nonumber\\
\Psi_u^j&=
\begin{bmatrix}
\tilde\psi^j\,(+,-)\\[2pt]
\psi_u^j\,(-,-)\\[2pt]
\hat\psi_d^j\,(+,-)\\
\end{bmatrix}
\,,&
\Psi_d^j&=
\begin{bmatrix}
\hat\psi^j\,(+,-)\\[2pt]
\hat\psi_u^j\,(+,-)\\[2pt]
\psi_d^j\,(-,-) \\
\end{bmatrix}\,,
\end{align}
and $\mathcal{S}^i=S^i(+,+)$,
where we decomposed the spinorial $SO(5)$ multiplets into $SU(2)_L$, up-type $SU(2)_R$ and down-type $SU(2)_R$ components, as required by the $SU(2)_L\times U(1)_R$ symmetry in the UV. The resulting zero modes correspond to the SM field content (including three right-handed neutrinos), $\psi_L^i$ and $\psi_{uR,dR}^i$ with quarks and leptons unified in $SU(4)$ representations, one vector-like representation $\chi_L$ and $\chi_R^\prime$, and three chiral SM singlets $S_L^i$. The SM-singlet fermions, the scalars $\Omega$ and $\Phi$, and the (global) $U(1)_\mathcal{S}$ symmetry are needed to give neutrinos masses via an inverse see-saw mechanism~\cite{Fuentes-Martin:2020pww}. The $\Omega$ field is also responsible for the spontaneous breaking of the UV gauge symmetry down to the 4321 symmetry, where the $U(1)$ factor in 4321 is the diagonal subgroup of $U(1)_l\times U(1)_R$. Furthermore, the $\Omega$ field can play the role of the GW scalar by appropriately choosing its brane-localized potentials and bulk masses. The scalar $\Sigma$ plays a key role in generating the light Yukawa couplings, as discussed below. Finally, the (global) $U(1)_\Psi$ symmetry~\bibnote{These global $U(1)$ symmetries could instead be gauge bulk symmetries broken at a deeper UV brane (like the Planck brane). In this case, anomaly cancellation requires UV-brane fermions with masses at the symmetry breaking scale.} is introduced to forbid baryon and lepton number violating higher-dimensional operators, already present in the Randall-Sundrum (RS) model~\cite{Randall:1999ee} (see e.g.~\cite{Huber:2000ie}). 

\subsection{Flavor hierarchies}

\begin{figure}[t]
\includegraphics[width=\columnwidth]{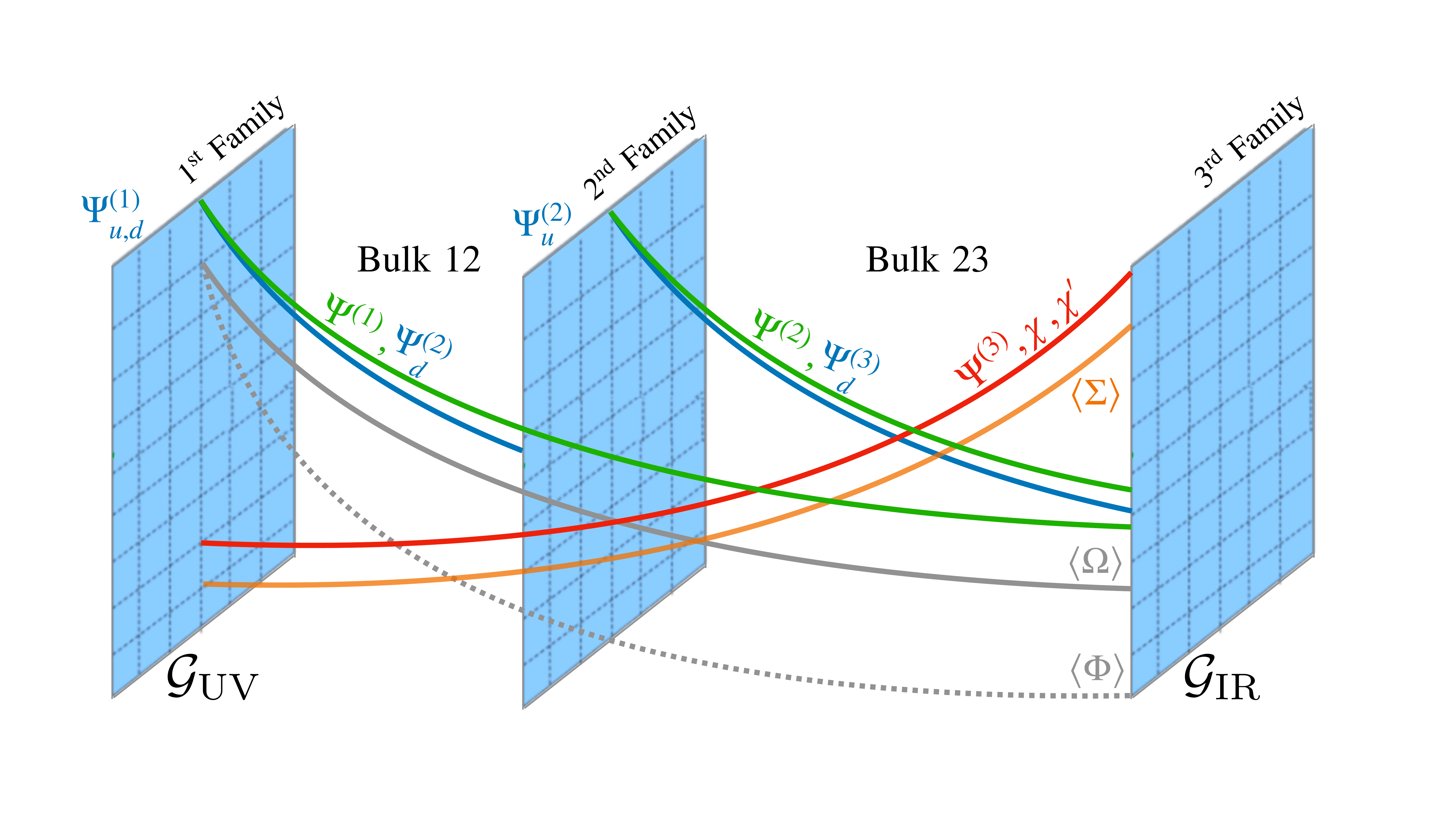}
\caption{Schematic structure of the extra dimension including fermion zero-mode and scalar VEV profiles.}
\label{fig:branes}
\end{figure}
As we discuss in what follows, {\it all} flavor hierarchies in the Yukawas and vector-like masses are explained with $\mathcal{O}(1)$ parameters by assuming the fermion localizations illustrated in Fig.~\ref{fig:branes}, with $kL\approx 10$ and $k\ell \approx 4$. While there is some freedom in the choice of fermion bulk masses that do not affect the solution to the flavor hierarchies, we take the following benchmark for concreteness (with $c_i\equiv M_i/k$ and $M_i$ the corresponding fermion mass)
\begin{align}\label{eq:cCoeff}
c_{\Psi^3}^{(1,2)}&=c_\mathcal{X}^{(1,2)}=c_{\Psi_d^3}^{(1)}=c_{\Psi^2}^{(1)}=c_{\Psi_u^2}^{(1)}=0\,,\;
c_{\mathcal{X}^\prime}^{(1,2)}=\eminus1/2\,,\nonumber\\
c_{\Psi^1}^{(1,2)}&=c_{\Psi^2}^{(2)}=\eminus c_{\Psi_d^3}^{(2)}=\eminus c_{\Psi_d^2}^{(1)}=1,\;
c_{\Psi_{u,d}^1}^{(1,2)},c_{\Psi_{u,d}^2}^{(2)}\leq\eminus2\,.
\end{align}
Even though the gauge symmetry in the 23-bulk is $SU(4)_h \times SU(3)_l$, for simplicity we choose $SU(4)$ symmetric bulk masses for this benchmark. Furthermore, we fix $\Lambda_{\rm IR}=8$~TeV such that $M_{15}=3.6$~TeV, which provides a good benchmark for the explanation of the $B$ anomalies~\cite{DiLuzio:2018zxy,Cornella:2019hct,Cornella:2021sby}.

\paragraph{Vector-like masses and fermion mixing.}

The following mass terms are added on the IR brane
\begin{align}
\mathcal{L}_{\rm IR}& \supset    \big(\bar{\mathcal{X}}_L \tilde M_\chi + \bar\Psi_L^3 \tilde M_{\Psi} + \bar \Psi_L^j \tilde m_\psi^j\big) \,\mathcal{P}_L \mathcal{X}^\prime_R\,,
\end{align}
in order to generate a vector-like mass among the zero modes. Here $\mathcal{P}_{L,R}$ is a projector into the $SU(2)_{L,R}$ components of the $SO(5)$ multiplets, and the IR masses decompose as $\tilde M_i=\mathrm{diag}(\tilde M_i^q,\tilde M_i^q,\tilde M_i^q,\tilde M_i^\ell)$ in $SU(4)$ space. These mass terms mix all the left-handed components of the zero modes, leaving the SM fermion content as massless chiral fields (plus the SM singlets needed for the inverse see-saw). The vector-like masses thus induced read
\begin{align}
\mathcal{L}&\supset-M_f\,\bar f_L\chi_R^\prime\,,\;\;
M_f=\Lambda_{\rm IR}\,\tilde M_f\, P\big(\{c_f^{(n)}\},\{c_{\mathcal{X}^\prime}^{(n)}\}\big)\,,
\label{VLmass}
\end{align}
where $P(\{c_1\},\{c_2\})$ is a fermion profile function (see~\ref{eq:ProfileFunction}).
The behavior of the profile function is such that, for the fermion profiles in~\eqref{eq:cCoeff}, we have 
\begin{align}
\Lambda_{\rm IR}\,P\big(\{c_{\Psi^j}^{(n)}\},\{c_{\mathcal{X}^\prime}^{(n)}\}\big)&\sim \frac{\Lambda_{\rm IR}}{\sqrt{kL}}\,e^{k(\ell_j-L)/2}\approx 2~\mathrm{TeV}\times V_{j3}\,,\nonumber\\
\Lambda_{\rm IR}\,P\big(\{c_{\Psi^3,\mathcal{X}}^{(n)}\},\{c_{\mathcal{X}^\prime}^{(n)}\}\big)&\sim \frac{\Lambda_{\rm IR}}{\sqrt{kL}}\approx 2~\mathrm{TeV}\,,
\end{align}
where $\ell_j$ is the position of the $j$-th brane, and $V_{ij}$ are CKM matrix elements. We thus obtain TeV-scale vector-like masses with a $U(2)$-like mixing structure for the assumed benchmark, hence reproducing the required conditions for a successful explanation of the $B$-meson anomalies~\cite{DiLuzio:2018zxy,Cornella:2019hct,Cornella:2021sby}.

\paragraph{Yukawa couplings.}

Yukawa couplings with the Higgs are generated via three distinct mechanisms, depending on the fermions, giving rise to decreasing effective interactions:

\medskip
{\bf I.} {\it Top Yukawa.} The fermion BCs in~\eqref{eq:FermionBCs} have been chosen such that, in the absence of IR masses, only the third-generation up-type Yukawa is generated:
\begin{align}
\mathcal{L}&\supset - y_u^{33}\,\bar\psi_L^3 H \psi_{uR}^3\,,\,
y_u^{33}=\frac{g_*}{2\sqrt{2}}\, P\big(\{c_{\Psi^3}^{(n)}\},\{c_{\Psi^3}^{(n)}\}\big)\,,
\end{align}
where $g_*$ is the $SO(5)$ KK coupling. In the absence of fermion mixing effects, $y_u^{33}$ becomes the top Yukawa, $y_t$. Since $P(\{c_{\Psi^3}^{(n)}\},\{c_{\Psi^3}^{(n)}\}\big)\lesssim1$, we infer the lower bound $g_{*}\geq 2\sqrt{2}\,y_t(\Lambda_{\rm IR})\approx2.2$. The relation between the EW gauge couplings and the $SO(5)$ KK coupling is  ($i=L,R$)
\begin{align}
g_i(\Lambda_{\rm IR})&=g_* /\sqrt{kL(1+r_{{\rm UV},i}^2+r_{{\rm IR},i}^2)}\,,
\end{align}
where $r_{{\rm UV (IR)},i}$ is the contribution from boundary kinetic terms for the corresponding gauge bosons at the UV (IR) brane evaluated at the IR scale~\cite{Csaki:2008zd}. Taking for simplicity $r_{{\rm IR},i}=r_{{\rm UV},i}\equiv r_i$, $kL=10$ and $g_*=2.5$, we find that the EW gauge couplings are reproduced for $r_L\approx0.5$ and $r_R\approx1.4$. 

\medskip
{\bf II.} {\it Other third-family Yukawas.} $b$ and $\tau$ Yukawas, as well as the leading mixing among the light families and the third generation, are generated only after introducing IR-brane mass-mixing terms. The relevant IR masses are
\begin{align}
\mathcal{L}_{\rm IR}&\supset \bar\Psi_L^3 \tilde M_{\Psi d }^L \mathcal{P}_L \Psi_{dR}^3+\bar{\mathcal{X}_L} (\tilde M_{\chi d }^L \mathcal{P}_L + \tilde M_{\chi d }^R \mathcal{P}_R)\Psi_{dR}^3\nonumber\\
&+\bar\Psi^j_L \tilde m_{\Psi j}^R \mathcal{P}_R \Psi_R^3+\bar\Psi^j_L (\tilde m_{dj}^L \mathcal{P}_L+\tilde m_{dj}^R \mathcal{P}_R) \Psi_{dR}^3\,,
\end{align}
where we ignored the mass term between the $SU(2)_R$ components of $\Psi^3$ and $\chi$, as well as those with $\Psi_{u,d}^{1,2}$, which have a minor phenomenological impact. These generate Yukawas between zero modes of the form
\begin{align}
y_{f1 f2}=\frac{g_*}{2\sqrt{2}}\, (\tilde M_{12}^L-\tilde M_{12}^R)\, P\big(\{c_{f_1}^{(n)}\},\{c_{f_2}^{(n)}\}\big)\,,
\end{align}
$f_1$ and $f_2$ denoting two generic 5D fermions, and $\tilde M_{12}^{L(R)}$ a generic IR mass between their $SU(2)_{L(R)}$ components.
As anticipated, the hierarchies between these and the top Yukawa are fully explained by appropriate fermion localizations. For the benchmark in~\eqref{eq:cCoeff}, we find 
\begin{align}
\begin{aligned}
P\big(\{c_{\Psi^j}^{(n)}\},\{c_{\Psi^3}^{(n)}\}\big)&\sim e^{k(\ell_j-L)/2} \approx V_{j3}\,,\\
P\big(\{c_{\Psi^3,\mathcal{X}}^{(n)}\},\{c_{\Psi_d^3}^{(n)}\}\big)&\sim  e^{k(\ell-L)/2}\approx y_b\,,\\
P\big(\{c_{\Psi^j}^{(n)}\},\{c_{\Psi_d^3}^{(n)}\}\big)&\sim e^{k(\ell+
\ell_j-2L)/2}\approx V_{j3}\,y_b \,,
\end{aligned}
\end{align}
where we used that $y_b\approx V_{23}$. Interestingly, we obtain a down-aligned limit (i.e.~vanishing light-heavy entries in the down sector) in the $SO(5)$ symmetric limit, where $M_i^L=M_i^R$. Due to the chosen fermion BCs, an analogous limit in the up sector is not possible. 

\medskip
{\bf III.} {\it Light-family Yukawas.} Due to the assumed strong UV localization of $\Psi_{u,d}^{1,2}$, light-family Yukawas are dominantly generated via their coupling with $\Sigma$. The relevant 5D proto-Yukawas are
\begin{align}
\mathcal{L}&\supset - Y_{u,d}^{ij}\,\bar\Psi^i\,\Sigma^a\,\Gamma^a\, P_R\Psi_{u,d}^j\,,
\end{align}
with $\Gamma^a$ ($a=1,\dots,5$) being the $SO(5)$ gamma matrices. These proto-Yukawas can be in the bulk as well as localized in the branes. The $\Sigma$ field decomposes under the EW symmetry as a Higgs $H^\prime$ and a singlet $S$, and acquires a vacuum expectation value (VEV) along the singlet direction with an IR-localized profile. This breaking of $SO(5)$ generates light-family Yukawas suppressed by the $\Sigma$ profile. Taking the $\Sigma$ bulk mass close to the Breitenlohner-Freedman stability
bound~\cite{Breitenlohner:1982jf,Breitenlohner:1982bm}, we find
\begin{align}\label{eq:LightYukawas}
y_{u,d}^{ij} &\approx  \frac{g_*}{2\sqrt{2}}\, \tilde Y_{u,d}^{ij}\,
\frac{\langle \Sigma_{\rm IR}\rangle}{\Lambda_{\rm IR}}\,e^{-k (L-\ell_j)}\nonumber  \\
&\quad\times e^{-k(c_i^{(1)}-\frac{1}{2})|y_i-\ell_j|}\,e^{k(c_j^{(1)}+\frac{1}{2})|y_j-\ell_j|}\,,
\end{align}
where $y_{i(j)}$ denotes the position of the brane where the left-handed (right-handed) field is dominantly localized, $c_{i(j)}^{(1)}$ is the left-handed (right-handed) 12-bulk mass in units of $k$, and $\tilde Y^{ij}_{u,d}$ are $\mathcal{O}(1)$ linear combinations of the proto-Yukawa couplings with coefficients depending on $k$, $\ell_i$, the fermion bulk masses, and the $\Sigma$ boundary masses (see~\ref{eq:tildeY}).

\subsection{Higgs potential and EW precision data}\label{subsec:HiggsPot}

The Higgs potential receives two types of contributions: i) a tree-level one resulting from the spontaneous breaking of the bulk gauge symmetry via $\Sigma$ and $\Omega$ VEVs, and ii) a loop-level one from a finite volume effect due to non-local operators generated by 5D loops stretching from one boundary to the other~\cite{HOSOTANI1983193}. In our model, the loop contribution comes dominantly from $\Psi^3$ and the EW gauge bosons. For small $h/f$, the Higgs potential is well approximated as
\begin{equation}
V(h) \approx \alpha(h) \cos\left(\frac{h}{f}\right) - \beta(h) \sin^2 \left(\frac{h}{f}\right) \,,
\end{equation}
where $f\approx 2\Lambda_{\rm IR}/g_{*}$ is the Higgs decay constant, $\alpha(h) \approx \alpha_\Omega + \alpha_{\Psi^3}(h)$ and $\beta(h) \approx \beta_\Sigma+\beta_{\rm EW} +\beta_{\Psi^3}(h)$. Using a holographic approach cross-checked by the spectral function method, we find the following expressions for the coefficients $\alpha_i$ and $\beta_i$~\bibnote{In the tree-level computation, we assume large $kL$ and neglect higher order terms in the boundary masses. For the loop via the spectral function, we include only $\Psi^3$ and EW gauge bosons, and we use approximate form factors as discussed in~\cite{Falkowski:2006vi}, which allow for an analytic computation.}
\begin{align}
\begin{aligned}
\alpha_{\Psi^3}(h) &\approx \frac{ 3N_c f^4 }{32\pi^2} \zeta(3)\, y_{t}^2 g_{*}^2 - 2\beta_{\Psi^3}(h)\,,
\\
\alpha_{\Omega}&\approx (\tilde M^R_{\Omega}-\tilde M^L_{\Omega})\,\Lambda_{\rm IR}^2 \langle \Omega_{\rm IR} \rangle^2\,, \\[5pt]
\beta_{\Psi^3}(h) &\approx  \frac{ N_c f^4 }{16\pi^2} \,  y_{t}^4 \left[\gamma + \log\frac{\Lambda_{\rm IR}^2}{m_{t}^2(h)} \right] \,, \\ 
\beta_{\rm EW} &\approx - \frac{9f^4}{512\pi^2}\, g_{*}^2\,  \zeta(3)  \left(3 g_{L}^2+g_{Y}^2\right) \,, \\
\beta_{\Sigma}&\approx \frac{1}{2}(\tilde M_{H^{\prime}}-\tilde M_{S}) \frac{\Lambda_{\rm IR}^2 }{(kL)^2} \langle \Sigma_{\rm IR} \rangle^2\,,
\end{aligned}
\end{align}
where $\gamma \approx 0.38$, $\zeta(3)\approx1.20$, $g_{L(Y)}$ is the $SU(2)_L$ ($U(1)_Y$) gauge coupling, $\langle \Omega_{\rm IR} \rangle$ and $\langle \Sigma_{\rm IR} \rangle$ are IR VEVs, $\tilde M^{L(R)}_{\Omega}$ are IR masses for the $SU(2)_{L(R)}$ components of $\Omega$, and $\tilde M_{H^{\prime},S}$ are UV masses for $\Sigma$ (all masses in units of $k$). Approximating $\alpha$ and $\beta$ as constants, the minimum of the potential and the Higgs mass are given by
\begin{align}
\cos(\langle h\rangle /f) &= -\frac{\alpha}{2\beta}\,\,,&
m_h^2\equiv 2\lambda \langle h\rangle^2 \approx \frac{2\beta\langle h\rangle^2}{f^4}\,.
\end{align}
As we can see, the loop contributions $\alpha_{\Psi^3},\beta_{\Psi^3},\beta_{\rm EW}$ are completely fixed once a value for $g_*$ (which also enters in the top Yukawa) is specified.
Interestingly, we find that the coefficients $\alpha_i$ and $\beta_i$ are of the right size such that the Higgs quartic comes out at the observed value for a natural choice of the undetermined tree-level parameters~\bibnote{We expect $\langle \Sigma_{\rm IR} \rangle \approx \mathcal{O}(\Lambda_{\rm IR})$. However, due to the localization of $\Omega$ in the UV, we can naturally have  $\langle \Omega_{\rm IR} \rangle \lesssim \Lambda_{\rm IR }$.}. While the $\alpha_i$ and $\beta_i$ are all of the same order, obtaining the required hierarchy between $\langle h\rangle$ and $f$ implies a tuning of the parameters (the so-called little hierarchy problem), which is at the per mille level for $\Lambda_{\rm IR}=8$~TeV and $g_*=2.5$. We have verified that the results of this simplified computation hold up to small corrections when treating the full loop potential numerically including all fields dominantly localized in the IR, as well as the relevant IR boundary masses.

Additionally, our theory predicts a tower of KK vector resonances which couple dominantly to the IR localized fields. The most dangerous of these states are those that mix with the SM EW gauge bosons, as they induce $g_*/g_L$ enhanced modifications to their couplings with third generation SM fermions. Resumming the KK tower, we find corrections of the form
\begin{equation}
\frac{\delta g_{Z \Psi^3 \Psi^3}}{g_{Z \Psi^3 \Psi^3}} \approx -0.3 \frac{m_Z^2}{M_{\rm KK}^2} \frac{g_*^2}{g_L^2} \approx - \frac{0.3}{4c_W^2} \frac{\langle h\rangle^2}{f^2} \,,
\label{eq:deltagOng}
\end{equation}
where $c_W$ is the cosine of the Weinberg angle and the pre-factor of $0.3$ comes dominantly from the Higgs and $\Psi^3$ profiles. The strongest bound comes from $Z\rightarrow \tau_L \tau_L$, leading to a constraint on~\eqref{eq:deltagOng} at the per-mille level~\cite{Efrati:2015eaa}, which is well satisfied for our benchmark point.

\section{Conclusions}\label{sec:conclusions}
We have presented a UV extension of the SM that addresses two of its long-standing open issues: the origin of flavor hierarchies and the stabilization of the Higgs sector, while, at the same time, explaining the observed anomalies in $B$ decays. A coherent solution to these three problems is obtained by embedding the SM into a warped 5D construction with three (flat) four-dimensional branes, where each SM family is quasi-localized. The 3-brane structure, which lets us associate flavor indices to well-defined positions in the extra dimension, is a crucial distinction from previous explanations of the flavor hierarchies in the context of warped extra dimensions~\cite{Grossman:1999ra,Nelson:2000sn,Agashe:2004cp,Csaki:2008zd}. This structure results in an approximate $U(2)^n$ flavor symmetry with leading breaking in the left-handed sector, which is necessary in order to evade the tight bounds on new physics from flavor-changing processes while simultaneously addressing the $B$ 
anomalies~\cite{Buttazzo:2017ixm,Barbieri:2015yvd}.

We emphasize that the explicit model analyzed here is part of a larger class of theories, based on the three fundamental points presented in the Introduction. A few building blocks, such as the choice of the IR-bulk and UV-brane symmetries, are motivated by observations. Other aspects (especially those related to UV dynamics) are less constrained and could be modified. The geometry itself is a minimal choice, and the validity of the construction could be extended up to the Planck scale by adding an additional UV brane, solving the large EW hierarchy problem as in the original RS model~\cite{Randall:1999ee}.

By construction, the TeV-scale phenomenology of this model is equivalent to that of 4321 models discussed in the recent literature~\cite{DiLuzio:2018zxy,Cornella:2019hct,Cornella:2021sby}. However, deviations are expected around $M_{\rm KK} \sim 10$~TeV due to the tower of KK states. A further striking signature, specific of the multi-scale (multi-brane) structure of the theory, is a multi-peaked stochastic gravitational wave signal potentially within reach of future experiments, originating from a series of phase transitions in the early universe~\cite{Greljo:2019xan}.

\vspace{1cm}

\subsection*{Acknowledgments}

This project has received funding from the European Research Council (ERC) under the European Union's Horizon 2020 research and innovation programme under grant agreement 833280 (FLAY), and by the Swiss National Science Foundation (SNF) under contract 200020\_204428. The work of JF has been supported by the Spanish Ministry of Science and Innovation (MCIN) and the European Union NextGenerationEU/PRTR under grant IJC2020-043549-I, by the MCIN grant PID2019-106087GB-C22, and by the Junta de Andaluc\'ia grants P18-FR-4314 (FEDER) and FQM101.

\begin{widetext}

\appendix

\section{General profile function}

The profile function introduced in the main text is given by
\begin{align}\label{eq:ProfileFunction}
P\big(\{c_L^{(n)}\},\{c_R^{(n)}\}\big)&\equiv p\big(\{c_L^{(n)}\}\big)\,p\big(\{-c_R^{(n)}\}\big)\,,
\end{align}
where, in the two bulk case with $k_1=k_2\equiv k$, we have
\begin{align}
p(\{c_1,c_2\})=\sqrt{\frac{(1-2c_1)(1-2c_2)}{(1-2c_1)-e^{2k(L-\ell)(c_2-1/2)}\left[(1-2c_2)\,e^{2k\ell(c_1-1/2)}+2(c_2-c_1)\right]}}\,,
\end{align}
with $\ell$ being the location of the intermediate brane. Some interesting cases are: $p\big(\{0,0\}\big)\approx 1$, $p\big(\{1/2,1/2\}\big)\approx \frac{1}{\sqrt{kL}}$, and $p\big(\{0,1\}\big)\approx \frac{1}{\sqrt{2}}\,e^{-k(L-\ell)/2}$, $p\big(\{1,1\}\big)\approx e^{-kL/2}$.

\section{Light-Yukawa couplings}

The parameters appearing in the light-Yukawa couplings formula~\eqref{eq:LightYukawas} are 
\begin{align}\label{eq:tildeY}
\tilde Y_{u,d}^{i2} =& 2\sqrt{2} \frac{\ell}{L}\,N_{y_i}(\{c^{(n)}_{i}\})N_{y_{c,s}}(\{-c^{(n)}_{c,s}\})\, \left(\frac{\sqrt{k}\,Y_{u,d}^{i2\,(1)}}{2-c^{(1)}_{i}+c^{(1)}_{c,s}}+\frac{\sqrt{k}\,Y_{u,d}^{i2\,(2)}}{-2+c^{(2)}_{i}-c^{(2)}_{c,s}}+\sqrt{k^3}\, Y_{u,d}^{\prime\prime\,i2}\right)+ \mathcal{O}\left(\frac{1}{kL}\right)\,,\nonumber\\
\tilde Y_{u,d}^{i1} =& \frac{2\sqrt{2}}{kL}\,N_{y_i}(\{c^{(n)}_{i}\})N_{y_{u,d}}(\{-c^{(n)}_{u,d}\})
\frac{1}{2\left(\tilde M_{H^{\prime}}+2\right)}
\left[\sqrt{k}\,Y_{u,d}^{i1(1)}\frac{6-c^{(1)}_{i}+c^{(1)}_{u,d}+2\tilde M_{H^{\prime}}}{\left(2-c^{(1)}_{i}+c^{(1)}_{u,d}\right)^2}-\sqrt{k^3}\, Y_{u,d}^{\prime\,i1}\right]+\mathcal{O}\left(\frac{1}{k^2L^2}\right)\,,
\end{align}
where $Y^{ij(1,2)}_{u,d}$ are the 12- and 23-bulk proto-Yukawas, $Y^{^{\prime}\,ij}_{u,d}$ ($Y^{^{\prime\prime}\,ij}_{u,d}$) are the first (second) brane proto-Yukawas, $c_{i}^{(1,2)}$ ($c_{u,d,c,s}^{(1,2)}$) are the left-handed (right-handed) fermion 5D masses in the 12- and 23-bulk in units of $k$, $y_i$ ($y_{u,d,c,s}$) is the position of the brane where the left-handed (right-handed) fermions is mainly localized, and
\begin{align}
N_y(\{c^{(n)}\})=
\begin{cases}
\sqrt{-1+2c^{(1)}}~~~~~~~~~~~~~~~~~\,{\rm if}~y=0 \\[5pt]
\sqrt{\frac{\left(-1+2c^{(1)}\right)\left(-1+2c^{(2)}\right)}{2\,(c^{(1)}-c^{(2)})}}~~~~~\,{\rm if}~y=\ell
\end{cases}\,.
\end{align}

\phantom{a}
\end{widetext}

\bibliographystyle{JHEP}
\bibliography{references}

\end{document}